\documentclass[sigconf]{acmart}

\setcopyright{rightsretained}
\acmYear{2024}
\copyrightyear{2024}
\acmConference[ASIA CCS '24]{ACM Asia Conference on Computer and Communications Security}{July 1--5, 2024}{Singapore, Singapore}
\acmBooktitle{ACM Asia Conference on Computer and Communications Security (ASIA CCS '24), July 1--5, 2024, Singapore, Singapore}
\acmDOI{10.1145/3634737.3637640}
\acmISBN{979-8-4007-0482-6/24/07}

\usepackage{etoolbox}
\makeatletter
\patchcmd{\maketitle}{\@copyrightspace}{}{}{}
\makeatother

\usepackage[nolist]{acronym}
\begin{acronym}[SIGMAR]
        \acro{CPS}{cyber-physical system}
        \acro{DoS}{Denial of Service}
        \acro{ICS}{Industrial Control System}
        \acro{OT}{Operational Technology}
        \acro{IDS}{intrusion detection system}
        \acro{IPS}{intrusion prevention system}
        \acro{MAC}{Message Authentication Code}
        \acro{IIoT}{Industrial Internet of Things}
        \acro{TEE}{Trusted Execution Environment}
\end{acronym}

\usepackage{amsfonts}
\usepackage{subcaption}
\usepackage{layouts}
\usepackage{hyperref}
\usepackage{enumitem}

\usepackage{xcolor}
\usepackage[detect-weight=true,detect-family=true]{siunitx}
\usepackage{xspace}

\usepackage{amsmath}

\usepackage{tikz}
\usetikzlibrary{positioning}
\usetikzlibrary{shapes.geometric}

\usepackage{multirow}

\usepackage{textgreek}

\definecolor{darkgrey}{RGB}{80,80,80}
\definecolor{lightgrey}{RGB}{170,170,170}

\definecolor{brown}{HTML}{a52a2a}
\definecolor{darkcyan}{HTML}{0a888a}

\usepackage{adjustbox}
\usepackage{booktabs}
\usepackage{wasysym}
\usepackage{tabularx} 
\usepackage{multirow}
\usepackage{arydshln} 
\usepackage{siunitx}

\newcommand{\bad}{
    \begin{tikzpicture}
        \node[draw,shape=circle,rotate=90,minimum width=2mm,fill=white, anchor=south, inner sep=0mm] (a) at (0,0) {};
    \end{tikzpicture}
}
\newcommand{\ok}{
    \begin{tikzpicture}
        \node[draw,shape=semicircle,rotate=90,minimum width=2mm,fill=black, anchor=south, inner sep=0mm] (a) at (0,0) {};
        \node[draw,shape=semicircle,rotate=270,minimum width=2mm,fill=white, inner sep=0mm,  yshift=0.3mm] (b) {};
    \end{tikzpicture}
}
\newcommand{\good}{
    \begin{tikzpicture}
        \node[draw,shape=circle,rotate=90,minimum width=2mm,fill=black, anchor=south, inner sep=0mm] (a) at (0,0) {};
    \end{tikzpicture}
}

\newcommand{\etal}{\textit{et~al.}\xspace}
\newcommand{\ie}{\textit{i.e.},\xspace}
\newcommand{\eg}{\textit{e.g.},\xspace}
\newcommand{\cf}{\textit{cf.}\xspace}

\newcommand{\wrt}{w.r.t.\xspace}

\newcommand{\code}[1]{\texttt{#1}}

\newcommand{\CUT}[1]{{\iffalse#1\fi}}

\newcommand{\mh}[1]{{\color{orange}{#1}}}

\usetikzlibrary{decorations.pathreplacing}

\newcommand*\circled[1]{\tikz[baseline=(char.base)]{
            \node[shape=circle,draw,inner sep=.5pt] (char) {#1};}}

\newcommand{\name}{\textsc{Madtls}\xspace}

\begin{document}

\title{\name: Fine-grained Middlebox-aware End-to-end Security for Industrial Communication }

\author{Eric Wagner}
  \email{eric.wagner@fkie.fraunhofer.de}
  \affiliation{%
    \institution{Fraunhofer FKIE}
    \country{}
  }
  \affiliation{%
    \institution{RWTH Aachen University}
    \country{}
  }

  \author{David Heye}
  \email{david.heye@rwth-aachen.de}
  \affiliation{%
	\institution{RWTH Aachen University}
    \country{}
  }
  \affiliation{%
    \institution{Fraunhofer FKIE}
    \country{}
  }

  \author{Martin Serror}
  \email{martin.serrror@fkie.fraunhofer.de}
  \affiliation{%
    \institution{Fraunhofer FKIE}
    \country{}
  }

  \author{Ike Kunze}
  \email{kunze@comsys.rwth-aachen.de}
  \affiliation{%
	\institution{RWTH Aachen University}
    \country{}
  }

  \author{Klaus Wehrle}
  \email{wehrle@comsys.rwth-aachen.de}
  \affiliation{%
	\institution{RWTH Aachen University}
    \country{}
  }

  \author{Martin Henze}
  \email{henze@spice.rwth-aachen.de}
  \affiliation{%
	\institution{RWTH Aachen University}
    \country{}
  }
  \affiliation{%
    \institution{Fraunhofer FKIE}
    \country{}
  }

\renewcommand{\shortauthors}{Wagner et al.}

\begin{abstract}
Industrial control systems increasingly rely on middlebox functionality such as intrusion detection or in-network processing.
However, traditional end-to-end security protocols interfere with the necessary access to in-flight data.
While recent work on middlebox-aware end-to-end security protocols for the traditional Internet promises to address the dilemma between end-to-end security guarantees and middleboxes, the current state-of-the-art lacks critical features for industrial communication.
Most importantly, industrial settings require fine-grained access control for middleboxes to truly operate in a least-privilege mode.
Likewise, advanced applications even require that middleboxes can inject specific messages (e.g., emergency shutdowns).
Meanwhile, industrial scenarios often expose tight latency and bandwidth constraints not found in the traditional Internet.
As the current state-of-the-art misses critical features, we propose Middlebox-aware DTLS (\name), a middlebox-aware end-to-end security protocol specifically tailored to the needs of industrial networks.
\name provides bit-level read and write access control of middleboxes to communicated data with minimal bandwidth and processing overhead, even on constrained hardware.

\end{abstract}




\maketitle

\section{Introduction}
\label{sec:introduction}

With the rise of the \ac{IIoT}, \acp{ICS} heavily rely on machine-to-machine communication between constrained devices to realize control of delicate physical processes~\cite{2016_luvisotto_ultra}.
Here, the timely and frequent exchange of short and predictable sensor and command messages is essential to ensure precise control over, \eg robot arms~\cite{2018_ruth_towards}.
However, this reliable data exchange is heavily threatened by a surge of attacks on the underlying industrial networks~\cite{2020_alladi_industrial}.
Due to a widespread lack of security measures in these networks, attackers can eavesdrop on  traffic or even manipulate it, potentially causing severe (physical) harm~\cite{osti_1505628}.
To thwart such attacks, the de-facto standard is end-to-end security, primarily in the form of TLS.

While TLS experiences wide-scale adoption in the traditional Internet, the prospects of deployments in industrial networks look rather grim.
One major roadblock in deploying end-to-end security is the widespread use of middleboxes that rely on deep packet inspection:
Traditional middlebox functionality, such as intrusion detection, requires access to sensor and actuator data to monitor industrial processes~\cite{2022_wolsing_ipal}, and the increasing interest in in-network computing likewise requires read and write access to in-flight data~\cite{2020_mai_network}.
Hence, middleboxes prevent the deployment of traditional end-to-end security, raising a serious security dilemma.

\begin{figure}
        \centering
        \includegraphics[trim={0 0.31cm 0 0},clip]{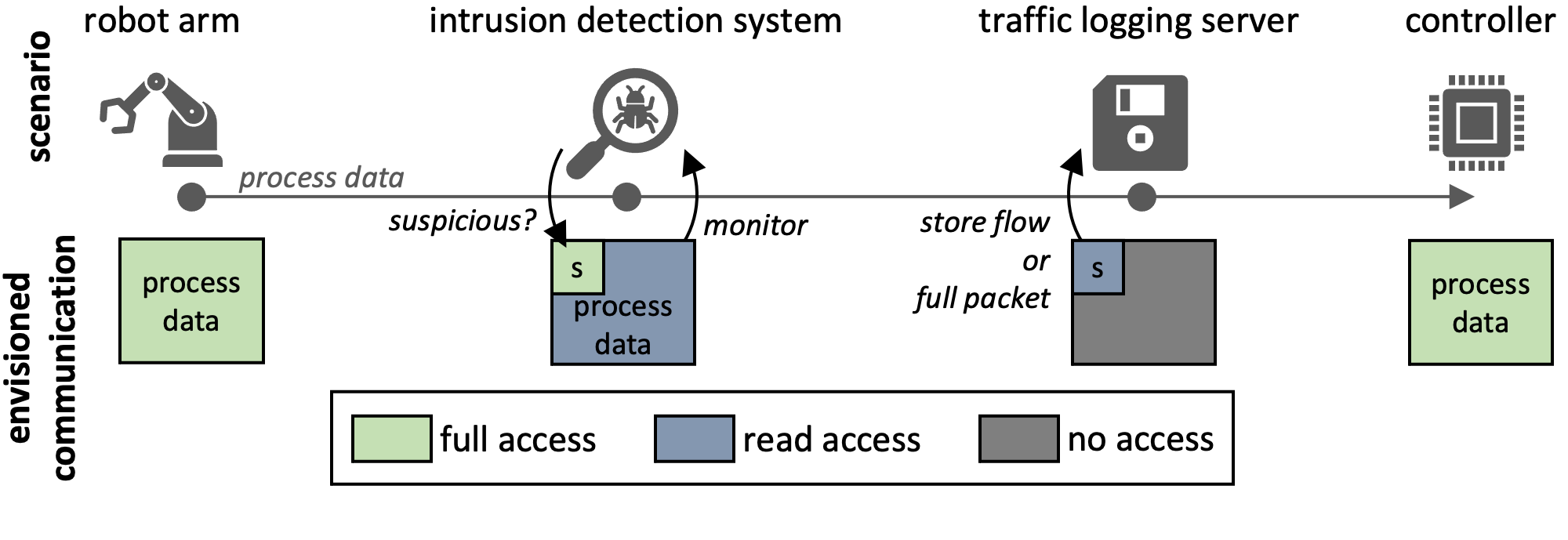}
        \caption{Middleboxes should operate in a least-privilege mode.
        For example, an industrial \ac{IDS} has read-only access to packets, and write access to a dedicated flag to mark suspicious traffic.
        A logging server can only read this flag.}
        \label{fig:example}
\end{figure}

To depict this issue, we consider an illustrative example of how interconnected \ac*{IIoT} communication may look in the future in Figure~\ref{fig:example}.
A controller operates a robot arm.
An \ac{IDS} and a traffic logging server monitor their communication.
The industrial \ac{IDS} analyzes traffic to detect anomalies and flags suspicious packets.
The traffic logging server only captures flow metadata but stores suspicious packets entirely for potential subsequent forensic investigations.

The na\"ive approach for middlebox access control while still providing security is to create separate secure point-to-point connections, one from robot arm to \ac{IDS}, one from \ac{IDS} to logging server, and a final one from logging server to controller. 
This approach, commonly referred to as SplitTLS~\cite{2015_naylor_mctls}, gives each middlebox full access to manipulate messages. 
Consequently, a compromised \ac{IDS} could move the robot arm unexpectedly and thus damage expensive equipment or cause physical harm.
Since middleboxes are typically placed at critical vantage points with access to large amounts of traffic, they then become especially attractive targets for attackers.
Therefore, it is of utmost importance to deploy middleboxes in a \emph{least-privilege mode}.
Considering the \ac{IDS} from our example, it should only have read access, except for writing to a single flag to mark suspicious packets.
In contrast, SplitTLS provides full insight and control over the entire communication channel to middleboxes.

To address these limitations, \emph{middlebox-awareness} has been proposed to adapt end-to-end security protocols to the reality of middleboxes in corporate networks and the Internet~\cite{2023_decarne_survey}.
A first branch of research relies on searchable encryption~\cite{2015_sherry_blindbox, 2016_lan_embark, 2017_canard_blindids, 2019_ning_privdpi} or zero-knowledge proofs~\cite{2022_grubbs_zero,2023_zhang_zombie} to perform a limited set of computations (\eg string matching) on encrypted data.
While these approaches may work for basic rule-based intrusion detection, they are too restrictive for most middleboxes in the industrial context.
Other proposals move middleboxes into \acp{TEE}~\cite{2017_han_sgx, 2018_poddar_safebricks, 2018_trach_shieldbox, 2019_duan_lightbox}.
While such approaches work great in theory, the secure implementation of the concept of \acp{TEE} is difficult in practice, as recent attacks have shown~\cite{2019_murdock_plundervolt, 2022_Borrello_AEPIC}.
Finally, a branch of research extends TLS to allow for the authorization of on-path middleboxes to read or write to a predetermined set of communicated data~\cite{2015_naylor_mctls, 2017_naylor_mbtls, 2019_lee_matls, 2019_li_metls,2023_ahn_mdtls,2023_fischlin_stealth}.
Still, operating on the granularity of complete messages or only supporting two access modes, these proposals do not provide the fine-grained access control necessary to operate middleboxes in a least-privilege mode.

While middlebox-aware end-to-end security thus offers an attractive solution for the security dilemma, current proposals do not address the unique challenges of industrial communication.
First, tight latency and bandwidth demands, paired with resource-constrained embedded hardware, require fast and efficient protocols~\cite{2016_luvisotto_ultra}.
Secondly, predictable and well-structured messages enable, but also require, fine-grained control down to the \emph{bit level} (instead of complete messages) over the access rights of each involved middlebox to constrain their privileges to a minimum.
Thereby, we can minimize the damage inflicted by potentially compromised middleboxes.
Thirdly, middleboxes may need to inject entirely new messages (\eg emergency shutdowns) into established communication channels, where the least-privilege principle must also be upheld.

This paper tackles these state-of-the-art limitations by proposing \textit{Middlebox-Aware DTLS} (\name).
In short, \name provides fine-grained access control to industrial communication via specialized cryptographic protocols.
We enable the transparent segmentation of messages to assign read and write access on a per-segment granularity.
Therefore, each segment is encrypted and authenticated individually.
Here, \name leverages a specifically tailored message authentication scheme to aggregate authentication data, thus conserving valuable bandwidth. 
Moreover, \name provides an efficient extension to the DTLS~1.2 handshake protocol to exchange the additional information required by the communicating endpoints and middleboxes.
While industrial networks may rely on reliable (\eg TCP) or lossy channels (\eg UDP, with a recent focus on wireless communication), we specifically target the more challenging domain where packets may be lost arbitrarily.
Still, the techniques designed in this paper are easily transferable to traditional TLS (over TCP) and other end-to-end protocols.

\textbf{Contributions.} To unlock the potential of middlebox-aware end-to-end security for resource-constrained industrial networks, we make the following contributions in this paper:
\begin{itemize}[noitemsep,topsep=2pt, after=\vspace{-2mm}]
        \item We provide a taxonomy of middlebox use cases in industrial communication and derive corresponding requirements for middlebox-aware end-to-end security~(Sections~\ref{sec:background}\,\&\,\ref{sec:motivation}).
        \item To meet these requirements, we design \name, which relies on specifically tailored cryptographic schemes to enable the distribution of fine-grained (down to the bit level) read and write access rights to middleboxes~(Section~\ref{sec:design}).
        \item To show the practical applicability and feasibility of our approach, we prototypically implement \name as DTLS~1.2 extension and show its competitive performance in realistic scenarios on representative hardware.
\end{itemize}

\section{Middleboxes in Industrial Networks}
\label{sec:background}

Middleboxes play an integral role in industrial networks for performance enhancements and intrusion detection~\cite{2023_liatifis_advancing}, but they also severely hinder the adoption of traditional end-to-end security.
In the following, we first give a brief background on industrial networks and why their unique properties hinder the use of traditional security protocols~(Section~\ref{sec:background:background}).
Afterwards, we provide an overview of middlebox functionalities in industrial networks, specifically focusing on the respective required access to communicated data~(Section~\ref{sec:background:usecases}).
Finally, we derive why the current state-of-the-art in middlebox-aware end-to-end security protocols cannot cope with the inherent requirements of industrial networks~(Section~\ref{sec:background:related_work}).

\subsection{Security Challenges in Industrial Networks}
\label{sec:background:background}

Industrial networks are dominated by machine-to-machine communication that keeps physical processes running safely.
Depending on the underlying process, this communication can be subject to harsh requirements \wrt latency and bandwidth~\cite{2012_galloway_ics}.
In particular, a plethora of sensors and actuators need to frequently exchange messages, hence per-message bandwidth overhead is especially expensive.
Meanwhile, devices must often be kept small and cheap, such that their processing resources are heavily constrained.

Communication has traditionally been facilitated via (reliable) cabled connections using different application layer and subordinate protocols.
Today, \ac{ICS} networks see a notable shift toward wireless and, thus, lossy communication, partly enabled by new protocols such as Sigfox or WirelessHART.
Therefore, any security solution cannot be tailored to a single protocol or communication medium but must be adaptable to reliable as well as lossy communication.

Concerning security, encrypted and authenticated end-to-end communication is still rarely seen within industrial networks~\cite{2022_dahlmanns_missed}, despite being commonplace in the traditional Internet.
The main reason for this lack of security is that, in the past, industrial networks were designed without security as their inherent physical separation from attackers seemingly sufficed.
As this assumption crumbles due to increased connectivity demands from industry~(\eg remote monitoring) and more advanced attacks, industrial networks become increasingly exposed to cyberattacks.
This vulnerability is demonstrated impressively through a rising numbers of attacks with, at times, detrimental consequences~\cite{2020_alladi_industrial}.

Addressing this lack of security and also accounting for the high demands in industrial networks, we observe an increasing interest in middlebox deployments~\cite{2020_mai_network}.
Security-wise, these middleboxes can, \eg realize deep packet inspection for intrusion detection.
Regarding performance, the full range of functionality extends to a diverse set of tasks including caching~\cite{2016_li_fast, 2021_gyorgyi_network}, data aggregation~\cite{2017_sapio_network}, fault detection~\cite{2019_saufi_challenges}, and in-network processing~\cite{2018_ruth_towards, 2019_rodriguez_network, 2020_cesen_towards}. 

\begin{figure}
        
        \begin{subfigure}{\columnwidth}
                \centering
                \includegraphics[trim={0 1.38cm 0 0},clip]{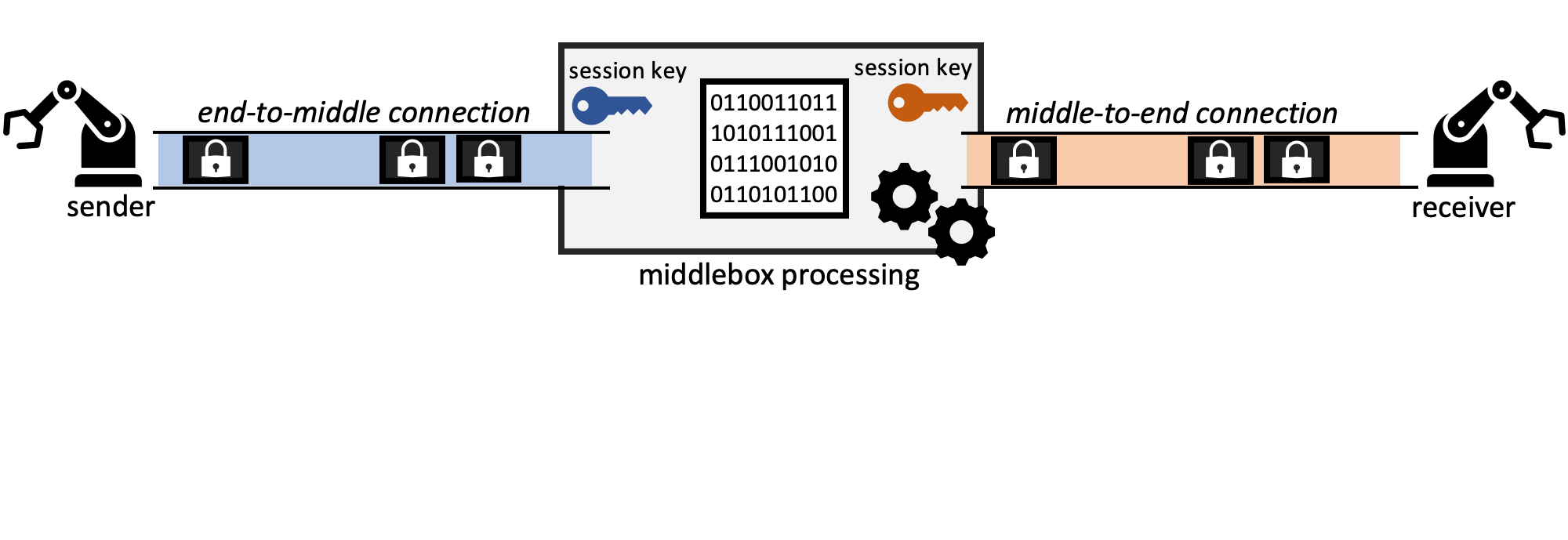}
                \caption{ SplitTLS solutions allow full access to middleboxes. }
                \label{fig:problem:problem} 
        \end{subfigure}
        \begin{subfigure}{\columnwidth}
                \centering
                \includegraphics[trim={0 1.38cm 0 0},clip]{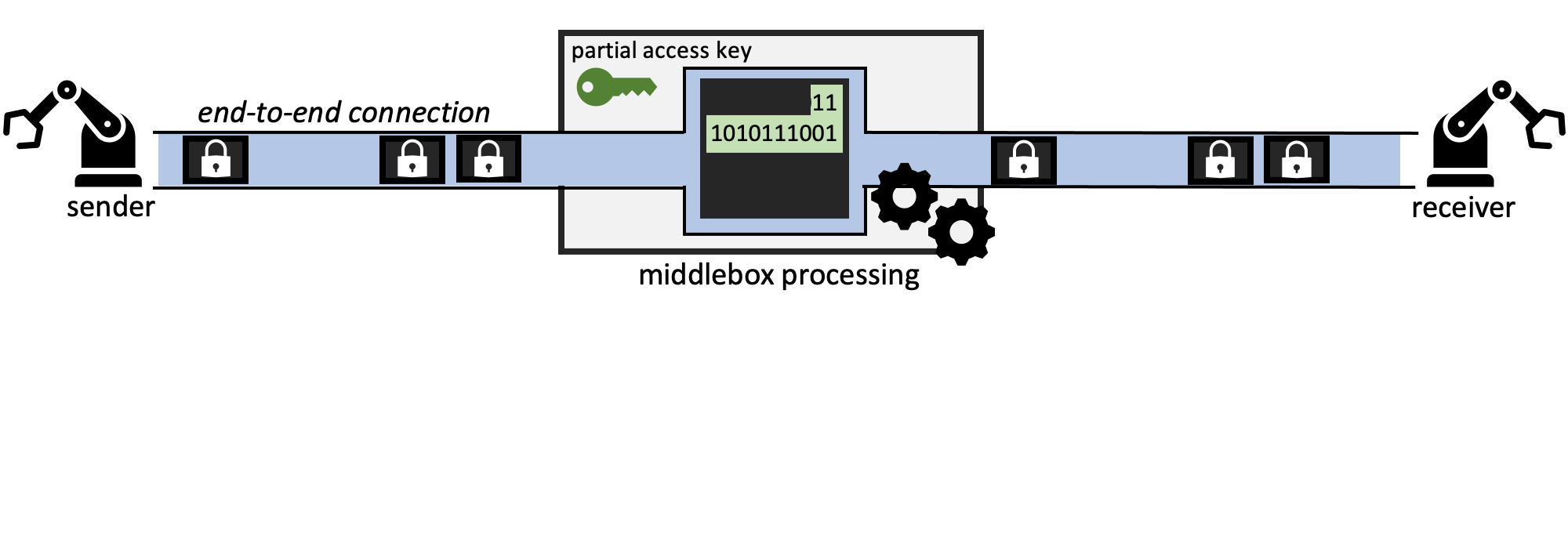}
                \caption{ A least-privilege access for middleboxes would be preferable.\vspace{-3mm}}
                \label{fig:problem:idea}
        \end{subfigure}
        \caption{While SplitTLS allows full access to middleboxes, achieving least-privilege access for middleboxes should be the goal for middlebox-aware security protocols.}
        \label{fig:problem}
\end{figure}

While these advancements alleviate specific limitations and \emph{some} security flaws within the \ac{IIoT}, they also hinder properly deploying end-to-end security protocols, as we demonstrate in Figure~\ref{fig:problem}.
Current industrial deployments of end-to-end security are mainly realized with TLS or DTLS and can, at most, provide a so-called SplitTLS~\cite{2015_naylor_mctls} solution, where middleboxes can freely access and modify any traffic that passes through them (\cf Figure~\ref{fig:problem:problem}).
However, middleboxes are usually deployed for a particular task, so an ideal security protocol would restrict read and write access to a minimum (\cf Figure~\ref{fig:problem:idea}).
Only then can the \ac{IIoT} operate in a least-privilege mode and minimize the damage of compromised middleboxes.


\begin{table*}[!htbp]
        \centering \footnotesize
        
        \newcolumntype{C}[1]{>{\centering}m{#1}}  
        
        \begin{tabularx}{\textwidth}{ 
                        m{2.6cm}
                        m{6mm}>{\raggedright\arraybackslash}
                        C{1.7cm}
                        C{3.5cm} 
                        C{.7cm}
                        C{.7cm}
                        m{.1cm}
                        C{.7cm}
                        C{.7cm}
                        m{.1cm}
                        C{.6cm}
                        C{.6cm}
                        C{.6cm}
                        m{.01cm}
                }
                
                &&&&\multicolumn{2}{c}{read-access}&&\multicolumn{2}{c}{write-access}\\[.1em] 
                \cline{5-6}\cline{8-9}
                \noalign{\smallskip} 
                \textbf{Publication}
                & \textbf{Year}
                & \textbf{Venue}
                & \textbf{Mechanism}
                & full
                & limited
                &
                & full
                & limited
                &
                & inject
                & drop
                & diff.
                &\\[.1em]\midrule
                BlindBox~\cite{2015_sherry_blindbox} & 2015 & SIGCOMM & searchable encryption  & \ok
                & \ok && \bad & \bad && \bad & \good & \bad & \\[.1em]
                Embark~\cite{2016_lan_embark} & 2016 & USENIX NSDI & searchable encryption & \ok & \ok && \bad & \bad && \bad & \good & \ok & \\[.1em]
                BlindIDS~\cite{2017_canard_blindids} & 2017 & ACM AsiaCCS & searchable encryption & \ok & \ok && \bad & \bad && \bad & \good & \bad & \\[.1em]
                PrivDPI~\cite{2019_ning_privdpi} & 2019 & ACM CCS & searchable encryption &  \ok & \ok && \bad & \bad && \bad & \good & \bad&\\[.1em]\midrule
                ZKMB~\cite{2022_grubbs_zero} & 2022 & USENIX Security & zero knowledge proofs & \ok & \ok && \bad & \bad && \bad & \ok & \good &\\[.1em]
                Zombie~\cite{2023_zhang_zombie} & 2023 & IEEE S\&P & zero knowledge proofs & \ok & \ok && \bad & \bad && \bad & \ok & \good &\\[.1em]\midrule
                SGX-Box~\cite{2017_han_sgx} & 2017 & APNet & trusted execution environment & \good & \bad && \good & \bad && \bad  & \ok & \bad & \\[.1em]
                Safebricks~\cite{2018_poddar_safebricks} & 2018 & USENIX NSDI & trusted execution environment & \good & \ok && \good & \ok && \bad  & \ok & \good & \\[.1em]
                Shieldbox~\cite{2018_trach_shieldbox} & 2018 & ACM SOSR & trusted execution environment  & \good & \bad && \good & \bad && \bad & \ok & \bad &\\[.1em]
                Lightbox~\cite{2019_duan_lightbox} & 2019 & ACM CCS &  trusted execution environment & \good & \bad && \good & \bad && \bad  & \ok & \bad & \\[.1em] \midrule
                mcTLS~\cite{2015_naylor_mctls} & 2015 & SIGCOMM & protocol extension & \good & \ok && \good & \ok && \bad & \bad & \good &\\[.1em]
                mbTLS~\cite{2017_naylor_mbtls} & 2017 & ACM CoNEXT & protocol extension & \bad & \bad && \good & \bad && \bad  & \bad & \good &\\[.1em]
                maTLS~\cite{2019_lee_matls} & 2019 & NDSS & protocol extension & \good & \ok && \good & \ok && \bad  & \bad & \good &\\[.1em]
                ME-TLS~\cite{2019_li_metls} & 2019 & IEEE IoTJ & protocol extension & \good & \ok && \good & \ok &&\bad  & \bad&  \good &\\[.1em]
                Stealth Key Exchange~\cite{2023_fischlin_stealth} & 2023 & ACM CCS & protocol extension & \good & \ok && \good & \ok &&\bad  & \bad & \ok &\\[.1em]
                mdTLS~\cite{2023_ahn_mdtls} & 2023 & ICISC & protocol extension & \good & \ok && \good & \ok &&\bad  & \bad & \good &\\[.1em]\midrule
                \textsc{Madtls} & 2024 & ACM AsiaCCS & protocol extension & \good & \good && \good & \good && \good & \ok & \good &\\[.1em]\midrule
                \multicolumn{13}{c}{ \hfill \good: yes \qquad \ok: partial \qquad \bad: no} 
                
        \end{tabularx}
        \caption{
                The current state-of-the-art on middlebox-aware security protocols cannot address  all requirements of industrial networks. 
                Searchable encryption or zero-knowledge proofs only provide limited functionality~(\eg string matching).
                Vulnerabilities within \acp{TEE} expose all approaches relying on them.
                Finally, extensions to the TLS protocol do not provide the fine-grained access control required to truly operate middleboxes in a least-privilege mode.
                Moreover, no proposal offers features to give middleboxes the ability to inject (a restricted set of) messages into the communication stream. 
        }
        \label{tab:related_work}
\end{table*}

\subsection{Diversity of Industrial Middlebox Use Cases}

\label{sec:background:usecases}

To understand the requirements for middlebox-aware end-to-end security in industrial networks, we need to understand the range of potential tasks.
Therefore, we categorize corresponding middlebox applications according to their required access rights.

\textbf{Read Access.}
Examples for read access include a middlebox that caches, \eg sensor readings to unburden the constrained end devices~\cite{2016_li_fast, 2021_gyorgyi_network}.
Similarly, industrial \acp{IDS} monitoring the physical process to detect suspicious activities only require read access to otherwise encrypted data for deep packet inspection~\cite{2022_wolsing_ipal}. 

\textbf{Limited Read Access.}
However, even for tasks such as intrusion detection or caching, middleboxes often only require partial insight into each message.
For example, most Snort~\cite{snort} rulesets for Modbus (a widespread industrial communication protocol) do not look beyond the function code field (indicating the type of a message, \eg request the state of an individual bit).
Even more sophisticated industrial \acp{IDS} only require partial read access to messages in some cases~\cite{2015_caselli_sequence, 2015_caselli_modeling, 2019_lin_timing}.
Moreover, other use cases only require partial read access by design:
For fault detection, middleboxes only need access to selected sensor readings from specific machinery to detect upcoming failures~\cite{2019_saufi_challenges}.
Similar access rights to only specific sensor readings are necessary for complex event detection, \eg the outbreak of a fire~\cite{2018_kohler_p4cep, 2020_mai_network, 2020_vestin_toward}.

\textbf{Write Access.}
Middlebox tasks that require \emph{full} write access to messages are rare, \eg when the middlebox translates between different application layer protocols~\cite{arrowhead, 2017_uddin_sdn}.
Instead, most middleboxes that alter messages only require limited write access.

\textbf{Limited Write Access.}
A typical example of limited write access in the industrial context and beyond is data compression, where a middlebox can, \eg base-delta encode timestamps for many different data sources~\cite{2018_pekhimenko_tersecades}.
Similarly, in-network aggregation or other map-reduce functionality can be executed by middleboxes that only have write access to the corresponding data fields they are reducing~\cite{2017_sapio_network}.
More industry-specific applications for restricted write access include, \eg the transformation of coordinates between reference frames~\cite{2021_kunze_investigating} or the insertion of precise timestamps into payload data~\cite{2020_kundel_p4sta}.
Furthermore, as seen in the example in Section~\ref{sec:introduction}, various middleboxes may take advantage of flagging individual packets, \eg to mark them as suspicious.

\textbf{Drop Messages.}
In the industrial context, it is often not desirable to directly drop messages due to irrevocable impact on the physical process controlled by the system.
Thus, industrial networks mostly only employ \acp{IDS} that flag suspicious traffic or alert the operator through other channels.
Still, for some use cases, a middlebox requires the ability to drop messages even in industrial networks, \eg to downsample sensor readings~\cite{2020_vestin_toward,Kunze:ISIE2021:SignalDetection}, carefully reduce traffic~\cite{2023_gyorgyi_adaptive}, or if a serious cyberattack is identified with high likelihood (\ie the benefit of likely preventing a high-impact attack outweighs the risk of blocking genuine traffic).


\textbf{Inject Messages.}
Finally, middleboxes may also require the ability to inject messages, most importantly when a middlebox directly responds to a request or event to reduce latency or traffic.
Complex event detection could, \eg identify critical conditions such as a fire~\cite{2018_kohler_p4cep}, that warrant the issuing of emergency stop messages~\cite{2020_cesen_towards}.
Other reasons to allow middleboxes to inject messages include responding caching servers~\cite{2016_li_fast, 2021_gyorgyi_network} or the enabling of low-latency and low-jitter control commands~\cite{2018_ruth_towards, 2019_rodriguez_network, 2020_cesen_towards}.

Overall, we see that middleboxes in the \ac{IIoT} cover a diverse set of functionalities that require different levels of access to a communication channel.
Most importantly, these functionalities are often specific to the \ac{IIoT} and must be considered when designing middlebox-aware end-to-end security.

\subsection{Prior Work on Middlebox-aware Security}
\label{sec:background:related_work}

Research on such \emph{middlebox-aware security} protocols goes back over a decade for the traditional Internet~\cite{2023_decarne_survey}, but we still see few deployments today.
Corresponding concerns can, among other things, be traced back to the Internet community’s end-to-end principle~\cite{RFC1958,RFC2775}, which is typically interpreted to imply that the functionality in the network should be kept minimal while end-hosts implement most, if not all, functionality.
Thus, especially in security contexts, the addition of middleboxes terminating end-to-end connections is often regarded as a slippery slope toward security and privacy loss on the Internet.
In contrast to the general Internet, limited domains~\cite{RFC8799} allow for more liberal solutions that can be tailored to the needs of the domain.
Industrial networks are one prominent example of such a limited domain, as they are typically under a single administrative control.
Additionally, all devices follow a common goal: ensuring the successful operation of the industrial process. 
Industrial networks and other limited domains thus represent a contrasting deployment scenario compared to the Internet, calling to revisit secure middlebox-aware communication specifically from the perspective of industrial communication.
Since current proposals for middlebox-aware end-to-end security protocols are designed for the general Internet, we now investigate to which extent existing proposals are suited for industrial deployments.
Table~\ref{tab:related_work} summarizes the results of our analysis.

An initial set of proposals attempts to realize middlebox functionality directly on encrypted traffic~\cite{2015_sherry_blindbox, 2016_lan_embark,2019_ning_privdpi, 2017_canard_blindids}.
Here, BlindBox~\cite{2015_sherry_blindbox} introduces the original idea but only enables the functionality to evaluate regular expressions on encrypted data.
Subsequent efforts extend this functionality~\cite{2016_lan_embark}, improve performance by reusing computations~\cite{2019_ning_privdpi}, or focus on specific use cases such as intrusion detection~\cite{2017_canard_blindids}.
However, searchable encryption enables only a limited set of computations on network traffic, does not support altering or injecting traffic, and brings unacceptable performance penalties to resource-constrained industrial devices.
A similar picture is drawn by the recent first proposal to employ zero-knowledge proofs provided by clients that attest the abidance to certain rules, \eg to prevent routing traffic to a blacklisted IP address~\cite{2022_grubbs_zero}.
Again, this approach has significant performance drawbacks, restricted functionality, and no support for altering or injecting traffic.

A fundamentally different idea is to encapsulate middlebox functionality into \acp{TEE} to shield them against malicious or compromised hosts~\cite{2017_han_sgx, 2018_poddar_safebricks, 2018_trach_shieldbox, 2019_duan_lightbox}.
These approaches, such as SGX-Box~\cite{2017_han_sgx} share TLS session keys with the \ac{TEE}. 
Within the \ac{TEE}, packets are entirely decrypted and can even be altered by the middlebox.
Lightbox~\cite{2019_duan_lightbox} further protects traffic metadata and enables stateful middleboxes.
Shieldbox~\cite{2018_trach_shieldbox} facilitates the implementation of middleboxes in Intel SGX via the \textsc{Click} framework~\cite{2000_kohler_click}.
Meanwhile, SafeBricks~\cite{2018_poddar_safebricks} restricts middleboxes to only partially access packets and improves the performances of chained middleboxes.
However, the access restriction of SafeBricks does not cryptographically protect against malicious middleboxes and relies on a correct realization of Rust's type system and the security of the \ac{TEE} itself.
Still, the limited memory of \acp{TEE} impedes the application of these proposals even if \acp{TEE} were available in industrial settings.
Furthermore, all protocols are designed for TLS, such that dropping individual messages implies that all future sequence numbers must be adapted.
More importantly, all these approaches assume a secure implementation of the \ac{TEE} primitive, which a plethora of recent attacks (\eg Plundervolt~\cite{2019_murdock_plundervolt} or \AE pic leak~\cite{2022_Borrello_AEPIC}) have shown to be difficult.

Finally, recent work investigates the direct integration of middleboxes into TLS sessions via protocol extensions~\cite{2015_naylor_mctls, 2019_lee_matls,2019_li_metls, 2017_naylor_mbtls,2023_ahn_mdtls,2023_fischlin_stealth}.
In mcTLS~\cite{2015_naylor_mctls}, each message is assigned a preconfigured context consisting of a unique set of middleboxes with read and/or write access.
The TLS record protocol header is then extended by a context identifier and two authentication tags, one for readers and one for writers.
Readers verify the reader tags, writers verify reader and writer tags, and endpoints verify all tags.
Thus, readers can verify that no third party modified the packet, writers know that any modification stem from writers, and endpoints additionally know whether the packet has been modified at all.
Still, mcTLS only provides coarse access control on a per-message level.
As an alternative to mcTLS, maTLS~\cite{2019_lee_matls} proposes to use different TLS sessions for middleboxes and append a modification log to each packet to track changes.
Despite performance improvements in mdTLS~\cite{2023_ahn_mdtls}, this approach introduces significant processing delay and bandwidth overhead.
In turn, ME-TLS~\cite{2019_li_metls} improves the handshake efficiency over that of mcTLS by providing authorized middleboxes with the necessary information to recover session keys from passively observed handshake messages.
Furthermore, mbTLS~\cite{2017_naylor_mbtls} enables the dynamic integration of middleboxes into a special TLS session where a new key is used for each middlebox to enforce path integrity, but without restricting data access.
Stealth key exchanges~\cite{2023_fischlin_stealth} take yet another route by exchanging a secondary encryption key in standard-conform TLS 1.3 communication that can be used to keep traffic encrypted and/or authenticated when sharing the primary session key with a middlebox.

Overall, even these closely related protocol proposals only provide access control on a per-message basis and often introduce significant overhead.
In general, the current state-of-the-art on middlebox-aware end-to-end security protocols cannot provide the fine-grained access control to packets necessary to operate middleboxes in a least-privilege mode.
Furthermore, no current proposal covers the case of middleboxes injecting traffic, as may be necessary within industrial applications to, \eg issue emergency stop commands~\cite{2020_cesen_towards} or reduce the latency of control tasks~\cite{2018_ruth_towards, 2019_rodriguez_network, 2020_cesen_towards}.
Consequently, we conclude that current middlebox-aware end-to-end security protocols are not suited for industrial networks.

\section{Threat Model \& Requirements}
\label{sec:motivation}

Despite extensive research on middlebox-aware end-to-end security, industrial networks can still not be efficiently equipped with such functionality.
Meanwhile, these networks offer a prime target for such approaches as established middlebox deployments often prevent other security measures.
In the following, we first establish the threat model against which middlebox-aware end-to-end security (in the industrial networks) must protect~(Section~\ref{sec:motivation:threat-model}).
Afterwards, we distill concrete requirements that must be fulfilled to enable the least-privilege operation of middleboxes~(Section~\ref{sec:motivation:requirements}).

\subsection{Threat Model}
\label{sec:motivation:threat-model}

We strive to prevent attacks aiming at unauthorized read or write access to communication channels in industrial networks.
To achieve this goal, we consider an attacker according to the Dolev-Yao threat model~\cite{1983_dolvey_yao}, \ie an attacker which has complete control over the entire network (but neither of the two endpoints of a communication channel).
Accordingly, the attacker can arbitrarily read, alter, reroute, inject, and drop packets.
Additionally, an attacker may arbitrarily compromise one or multiple middleboxes which access the communication channel.
Within our threat model, the attack must be constrained from extending their control over a communication session beyond the minimum access requirements any compromised middleboxes have over that session.
Furthermore, an attack must be prevented from altering middleboxes' order or skipping some middleboxes entirely, as this may lead to incorrect data being processed or ineffective intrusion detection.
\ac{DoS} attacks and side-channels attacks are out of scope in this paper as these kinds of attacks affect all security protocols. 

\subsection{Requirements}
\label{sec:motivation:requirements}

Middlebox-aware security protocols should provide the same security guarantees as end-to-end protocols towards outsiders, \ie entities not part of the communication.
Concretely, any communication session should authenticate the communication endpoints, provide data secrecy, and ensure data integrity.
Beyond these inherited requirements, middlebox-aware end-to-end security protocols (for industrial networks) must fulfill additional requirements regarding the integration of middleboxes into communication sessions.

\textit{\textbf{Explicit Middlebox Authentication.}}
The authentication of endpoints in end-to-end security must be extended to all middleboxes with read or write access to any message.
Thus, both endpoints must explicitly acknowledge and verify all middleboxes involved in the communication and the privileges they got assigned.

\textit{\textbf{Least Privilege Read and Write Access.}}
Middleboxes are often located at critical vantage points to process as much traffic as possible.
This makes middleboxes a particularly attractive target for attacks, while they often fulfill dedicated tasks which require read and/or write access to specific parts of messages.
Consequently, middleboxes should operate in a least-privilege mode where they are restricted to exactly those access rights that are inevitable to fulfill their task.
The selection of applications from Section~\ref{sec:background:usecases} shows that this access might have to be restricted to bit-wise read and/or write access to specific fields in a message.

\textit{\textbf{Least Privilege Traffic Injection.}}
Some middlebox tasks need to inject control commands for low-latency control or emergency shutdowns~(Section~\ref{sec:background:usecases}).
However, such abilities must not be accompanied with full control over the communication channels, \ie a middlebox should not be able to inject arbitrary traffic.
Instead, privileges to inject traffic must again be restricted to the minimum required for correct functionality, \eg to only inject messages with specific Modbus function codes.

\textit{\textbf{Path Integrity.}}
Generally, the order in which middleboxes process a message is important.
A middlebox performing complex computation may, \eg need to be placed behind a filtering middlebox to be able to keep up at line-rate.
Therefore, an attacker should not be able to change the processing order of middleboxes, or worse, skip certain middleboxes (\eg an \ac{IDS}) entirely.
To prevent such attacks, a middlebox-aware security protocol should enforce path integrity, \ie a message's correct verification depends on it passing all intended middleboxes in the right order.

\textit{\textbf{Accounting for Resource Constraints.}}
Middlebox-aware end-to-end security for industrial networks has to account for resource-constrained devices and networks.
More specifically, adequate latencies must be ensured even with limited processing power.
Additionally, any per-message overhead for short messages should be minimized to conserve bandwidth.

Our requirements show similarities with the traditional Internet~(\eg~path integrity~\cite{2017_naylor_mbtls}), but also a range of challenges unique to the \ac{IIoT}~(\eg least-privilege traffic injection).
As the current state-of-the-art cannot fulfill all these requirements, we propose a middlebox-aware security protocol tailored to the \ac{IIoT}.

\begin{figure}
        \includegraphics[width=\columnwidth]{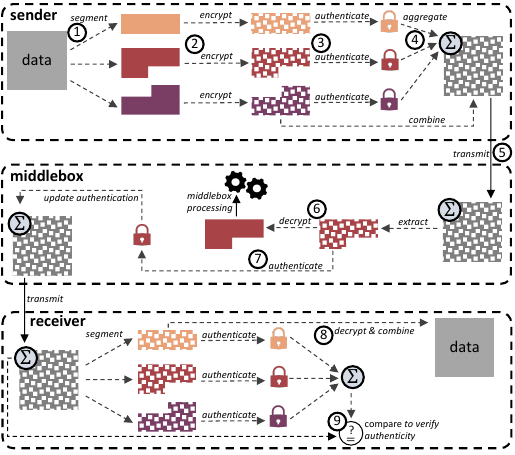}
        \caption{
                The core idea behind \name is to segment messages according to predetermined templates, which allows to encrypt and authenticate each segment individually.
                To save bandwidth, segment authentication is only verified by the receiver and selected middleboxes.
                Therefore, each middlebox updates the authentication tag, such that later verification of this tag ensures path integrity and data integrity at all times.
                }
        \label{fig:overview}
\end{figure}

\section{High-level Design of \name}
\label{sec:design}

To enable middlebox-aware end-to-end security in industrial networks, we propose Middlebox-Aware DTLS (short: \name).
We choose to integrate \name as extension to DTLS as it represents the bigger challenge compared to TLS, since any message can be lost during its transmission.
However, the integration into, \eg TLS 1.3 should be possible without significant changes.
\name is designed to fulfill the requirements for industrial networks outlined in Section~\ref{sec:motivation:requirements}.
Still, \name can also be advantageous in other scenarios such as data center networks.

The main idea underlying \name is to partition messages into segments, which are then individually encrypted and authenticated such that middleboxes can only read or write to a specific subset of those segments. 
We illustrate the entire process of securing a message before transmission, over the partial access by a middlebox, until the final reception of the message at its destination in Figure~\ref{fig:overview}.
First, \circled{1} the sender partitions a message into non-overlapping segments such that a specific set of access rights for each middlebox can be assigned to each segment.
The resulting segments are then \circled{2} separately encrypted (\eg with AES in counter mode), and \circled{3} an authentication tag is computed for each segment.
Notably, the authentication scheme for individual segments is designed such that \circled{4} all authentication tags for individual segments can be aggregated to save valuable bandwidth.
The encrypted segments and the aggregated authentication tag are then \circled{5} transmitted to their destination and intercepted by a middlebox.

In the example in Figure~\ref{fig:overview}, the middlebox has read access to the segment in the middle (red).
After \circled{6} decrypting this segment, it can process the contained data to realize its middlebox functionality.
Meanwhile, the middlebox also \circled{7} updates the aggregated authentication tag such that the final receiver can retrace whether this middlebox received the correct data.
Thus, the receiver can verify that no attacker manipulated data before a middlebox processes it, just to revert these changes afterwards.

The final receiver \circled{8} decrypts all segments and \circled{9} computes an authentication tag over the received data to compare it to the transmitted tag.
If both tags match, the integrity of the transmitted data is proven: 
The receiver \emph{and all middleboxes} received data that has only been modified by middleboxes that have been explicitly allowed to make these changes.
After discussing \name on a high level, we now dive into the details of the record layer protocol.

\section{The \name Record Protocol}
\label{sec:record_protocol}

To introduce the idea of middlebox-awareness to industrial networks, we extend the DTLS~1.2 protocol. 
We start with a summary of the DTLS~1.2 record layer layouts in Section~\ref{sec:design:dtlsbackground}.
Then, in Section~\ref{sec:design:header}, we discuss how we extend this layout to support limited read and write access for middleboxes.
In Section~\ref{sec:design:enc} and~\ref{sec:design:auth}, we then discuss how encryption and integrity protection are handled.
Finally, we discuss extensions to realize integrity verification by middleboxes~(Section~\ref{sec:design:verify}) and limited data injection~(Section~\ref{sec:design:inject}).

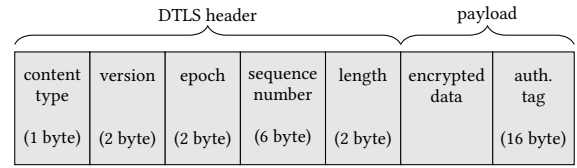
\begin{figure}[t]
	\centering
	\tikzstyle{rect}=[rectangle,minimum width=1cm,minimum height=1.5cm,draw=black,outer sep=0pt, align=center, font=\footnotesize]
	\tikzstyle{base rect}=[rect,minimum width=1cm,outer sep=0pt]
	\tikzstyle{text rect}=[rectangle,minimum width=1cm,align=center,outer sep=0pt]
	\tikzstyle{multipart rect}=[rectangle split,rectangle split horizontal,rectangle split parts = 2,inner xsep = 0.0mm,outer sep=0pt,
	draw=black,align=center,minimum width=2cm,minimum height=1cm]
	\begin{tikzpicture}
			
			\node (ct) [base rect,fill=gray!20] {content\\type\\\hfill\\(1 byte)};
			\node (ver) [rect,fill=gray!20,right=0.0cm of ct.east] {version\\\\\hfill\\(2 byte)};
			\node (epoch) [rect,fill=gray!20,right=0.0cm of ver.east] {epoch\\\\\hfill\\(2 byte)};
			\node (seq) [rect,fill=gray!20,right=0.0cm of epoch.east] {sequence\\number\\\hfill\\(6 byte)};
			\node (length) [rect,fill=gray!20,right=0.0cm of seq.east] {length\\\\\hfill\\(2 byte)};
			
			\node (data) [rect,fill=gray!20,right=0.0cm of length.east] { encrypted\\data \\\hfill\\\phantom{(2 byte)}};
			\node (tag) [rect,fill=gray!20,right=0.0cm of data.east] { auth. \\tag\\\hfill\\(16 byte)};
			
			\draw [decorate,decoration={brace,amplitude=7pt}]
			([yshift=1mm]ct.north west) -- ([yshift=1mm]data.north west) node [black,midway,yshift=4mm]{\footnotesize DTLS header};
			\draw [decorate,decoration={brace,amplitude=7pt}]
			([yshift=1mm]data.north west) -- ([yshift=1mm]tag.north east) node [black,midway,yshift=4mm]
			{\footnotesize payload};
			
	\end{tikzpicture}
	
	\caption{The DTLS 1.2 header format realizes a concise message format that can easily be extended to support new functionality through the addition of content types.}
	\label{fig:record-protocol}
\end{figure}

\subsection{Background: DTLS Record Layer}
\label{sec:design:dtlsbackground}

We first describe the DTLS record layer in Figure~\ref{fig:record-protocol}.
The \code{content type} is used to identify the type of messages (\eg DTLS handshake messages are represented by the type \texttt{0x16}).
New content types can be added to extend the functionality of DTLS.
The \code{version} field indicates the protocol version (\eg 1.2).
Then, DTLS uses two fields that combine to a nonce to prevent replay attacks.
The \code{epoch} tracks cipher suit changes and a zero-initialized \code{sequence number} increments with each message. 
The \code{length} field then indicates the length of the DTLS record layer payload, which is appended afterwards.
To this end, the payload is encrypted and an authenticated tag is appended according to the chosen cipher suite. 
In summary, DTLS realizes a concise message format that can easily be extended to new functionality through new content types.

\subsection{\name{}' Record Layer Header Structure}
\label{sec:design:header}

To realize \name{}, we extend the DTLS~1.2 header by three new content types (\texttt{0x1D}, \texttt{0x1E}, and \texttt{0x1F}).
One of those new content types (\texttt{0x1E}) represents the \name record layer.
The remaining header is extended by a, usually 1-byte long, \texttt{segmentation info} field that defines the encrypted payload's segmentation.

\begin{figure}
	\includegraphics[width=\columnwidth]{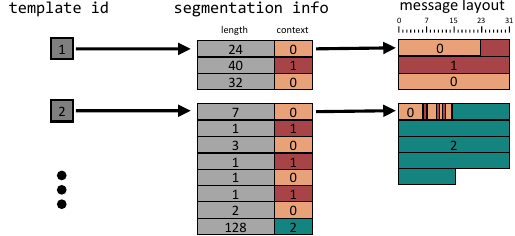}
	\caption{The \texttt{segmentation info} describes the layout of a packet and can either be explicitly transmitted, or, for repeating patterns, the mapping between a 1-byte \texttt{template id} and it can be communicated during the handshake.}
	\label{fig:template}
\end{figure}

This \texttt{segmentation info} field starts with two 1-bit flags: 
The \texttt{m}-flag, which is set to enable middlebox authentication (discussed in Section~\ref{sec:design:verify}), and the \texttt{l}-flag, which is set if the layout of the data is explicitly indicated.
If the \texttt{l}-flag is not set, the remaining 6 bits form the \texttt{template id} that maps to one of up to 64 pre-exchanged segmentation infos as shown in Figure~\ref{fig:template} and explained below.

Conceptually, a plaintext \name message is divided into $n$ segments, addressed as $S[0]$ to $S[n-1]$.
Each segment $S[\cdot]$ is assigned a context that defines which middleboxes have read or write access to that segment.
The segmentation of a message is then indicated indirectly through the \texttt{template id} or by explicitly including the \texttt{segmentation info} in the packet header.
The \texttt{segmentation info} encodes the layout where alternating fields indicate the bitlength of segments and their assigned context.
Currently, variable-length fields result in entire segmentation info specifications, but a compact formatting for such cases could be imagined in the future.

As in DTLS, the header precedes the payload that contains the encrypted message and the integrity-protecting authentication tag that is verified by the final receiver of a message.
Crucially, \name' authentication tag is updated by middleboxes to ensure data consistency and path integrity. 
In the following, we discuss the encryption and authentication scheme employed by \name in more detail.


\subsection{Segment Encryption}
\label{sec:design:enc}

We start by discussing the encryption scheme used by \name, before diving into the more important (for industrial networks) and challenging aspect of integrity protection.
Note that encryption is not mandatory in \name, which may be useful for some (industrial) applications that only care about integrity-protection.
When using encryption, the use of a single key does obviously not allow for limited read access.
Therefore, each segment $S[\cdot]$ is assigned a context $c$, which is associated with a unique encryption key $k_c^{enc}$.
Only those middleboxes that should read a specific context are then provided with the corresponding key.
These keys are distributed during the handshake, as we will discuss in Section~\ref{sec:handshake}.

To avoid unnecessary overhead when segments are shorter than a multiple of the block sizes of the cipher (\eg \SI{16}{bytes} for AES), \name can take advantage of cipher streams as realized, \eg by AES in counter mode.
This approach ensures that messages do not expand through encryption of individual segments.
Each segment is then separately encrypted with the key corresponding to its context, \ie $C[i] = enc_{k_c^{enc}}(S[i])$, where $C[i]$ is the encrypted plaintext segment $S[i]$.
Middleboxes and the final receiver derive either from the \emph{template id} or from the \textit{segmentation info} field which segments they have access to, where those are located, and which keys they must use to decrypt which segment. 

\subsection{Compact Authentication Scheme}
\label{sec:design:auth}

While \name' encryption scheme is straightforward, the similarly trivial approach towards integrity protection is impossible. 
\name's authentication scheme must differentiate individually between read and write access to all segments.
Consequently, we cannot transfer approaches such as mcTLS's three authentication tags~\cite{2015_naylor_mctls} to a setting with significantly more fine-granular access control, as its authentication scheme introduces three 16-byte tags for each context and does not protect path integrity.


Therefore, we design a custom authentication scheme for \name{} that ensures compactness, path integrity, and high performance.
To achieve these goals, \name takes advantage of authentication tag aggregation~\cite{2008_katz_aggregate} to combine multiple tags into a single tag without loss of security for a verifier that is able to verify each of the aggregated tags individually.
While tag aggregation has been explored previously to achieve path integrity~\cite{2023_esiner_message}, \name's authentication scheme only needs to transmit a single tag even if messages are modified or divided into multiple contexts.
This design naturally favors the verification of data and path integrity by the final receiver who has access to all contexts to verify all tags.
In case the receiver notices that a packet has been manipulated on its path, it can alert the concerned middleboxes or an operator.
For now, we focus on this case and describe the design of \name's single authentication tag in the following.
We later revisit this limitation in Section~\ref{sec:design:verify} and see how \name supports the efficient creation of partial authentication tags that middleboxes with limited data access can still verify.

The gist of our authentication scheme is that each node that is authorized to read or write data manipulates the authentication tag in a deterministic way such that the final tag is correct iff no unauthorized manipulation has taken place during message transmission.
The manipulation of the tag by each entity with access rights, even those only allowed to read, ensures that no entity receives manipulated data that is subsequently changed back without this being noticeable by the final receiver.
In the following, we describe \name' authentication scheme in detail, starting with the key setup. 
Afterwards, we learn how the sender computes an initial authentication tag and how this tag is subsequently updated by middleboxes according to their access rights.

\begin{figure}
        \includegraphics[width=\columnwidth]{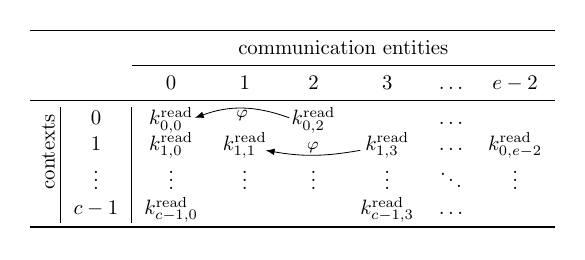}
        \vspace{-10mm}
        \caption{ \name assigns read keys to communication entities according to their access rights for each context. The same key assignment is conducted for write keys.}
        \label{fig:key-matrix}
\end{figure}

We have $e$ communication entities and $c$ contexts in a \name session.
Each segment $S[\cdot]$ is mapped to a context that uniquely describes a set of access rights for each middlebox.
Entities $0$ and $e-1$ are the sender and receiver, respectively.
All other entities ($1$ to $e-2$) are middleboxes with read and/or write access to a subset of all segments.
The corresponding symmetric keys for access to the $i$-th context for the $j$-th entity are denoted as $k_{i,j}$ as shown in Figure~\ref{fig:key-matrix}.
For each index, there exist up to two keys, one for read access ($k^{\text{read}}_{i,j}$) and one for write access ($k^{\text{write}}_{i,j}$).
As shown in the example of Figure~\ref{fig:key-matrix}, Entity~3 has read access to Context~1 such that $k^{\text{read}}_{1,3}$ exists.
As Entity~2 has no read access to this Context, $k^{\text{read}}_{1,2}$ does not exist.
The sender has read and write access to all contexts of a message.
Thus, all keys $k_{\_,0}$ always exist.
Meanwhile, no key $k_{\_,e-1}$ exists as the receiver does not need to authenticate the message to another entity.
To update authentication tags for the next entity, the $k$-th communication entity has access to all keys $k_{\_,k}$ that exist. 
To update (and verify) tags, each entity additionally knows the \emph{previous} existing key for all contexts it has access to.
We denote this previous key as  $\varphi (k_{\_,k})$.
In the example from Figure~\ref{fig:key-matrix}, $\varphi (k_{0,2}^\text{read}) =  k_{0,0}^\text{read}$ and $\varphi (k_{1,3}^\text{read}) =  k_{1,1}^\text{read}$. 

In the following, we explain how the sender uses the keys $k_{\_,0}$ to compute the initial authentication tag.
Each entity subsequently updates the aggregated authentication tags by removing the old {partial segment tag} (\ie the authentication tag computed only over a segment with the corresponding read or write key) and adding a new tag for each accessed context.
As each update requires access to a unique set of keys that only that specific middlebox knows, it cannot be impersonated by another entity.
An in-depth discussion of the soundness and security of \name' authentication scheme can be found in the Appendix.

The initial tag is computed as follows, where $\delta(\cdot)$ maps the segment index $i$ to the corresponding context: 
\begin{equation*}
  t = \bigoplus_{0\leq i<n} \left(  \sigma_{k^{\text{read}}_{\delta(i),0}}(C[i]) \oplus \sigma_{k^{\text{write}}_{\delta(i),0}}(C[i]) \right) 
\end{equation*}
Each segment is authenticated ($\sigma$ represents a classical message authentication algorithm such as HMAC-SHA256) twice, with the corresponding reading and writing keys, respectively.
All computed tags (\ie reading and writing tags for all segments) are then XOR-ed together, which does not reduce their security~\cite{1995_bellare_xor}.
A verifier now needs all keys that were used to compute the individual tags to verify the aggregated tag. 
This aggregated tag is then appended to the message and transmitted to the first middlebox with limited read or write access to the message.

All middleboxes alter this tag according to their access rights in a deterministic way.
A middlebox $j$ that has read access to the segments $\mathbb{S}^\text{read}_{j}$ updates the tag of the message as follows:
\begin{equation*}
        t \overset{\oplus}{=} \bigoplus_{i\in \mathbb{S}^\text{read}_{j} } \left(  
        \sigma_{ \varphi( k^{\text{read}}_{\delta(i),j}    )}(C[i]) 
        \oplus \sigma_{k^{\text{read}}_{\delta(i),j}}(C[i])  \right) 
\end{equation*}
For each segment $i$ in $\mathbb{S}^\text{read}_{j}$, the first part of the equation removes tags from the last entity that had read access to that message segment. 
This removal works exactly when the segment $C[i]$ has not been changed between the two readers, as only then do the old partial segment tag and the newly computed partial segment tag cancel out.
The second part of the equation then computes and integrates a new segment tag with the new key $k^{\text{read}}_{\delta(i),j}$ not known to the previous middlebox. 

Besides read access, some middleboxes may also have write access to certain segments $\mathbb{S}^\text{write}_{j}$.
Here, we assume that write access implies read access.
Formally, this means that $\mathbb{S}^\text{read}_{j}$ $\cap$ $\mathbb{S}^\text{write}_{j}$ = $\emptyset$, \ie write access to a context cannot be combined with explicit read access.
Here, the procedure to amend the authentication tag is similar. 
First, the old \emph{reading and writing} tags are removed before they are replaced by the new tag computed over the changed segment data $C'[\cdot]$.
Formally, this procedure looks as follows:
\begin{align*}
                t \overset{\oplus}{=} \bigoplus_{i\in \mathbb{S}^\text{write}_{j} } 
                \Bigl( & \sigma_{ \varphi( k^{\text{read}}_{\delta(i),j}    )}(C[i])
                \oplus \sigma_{k^{\text{read}}_{\delta(i),j}}(C'[i]) \oplus \phantom{} \\
                & \sigma_{\varphi(k^{\text{write}}_{\delta(i),j})}(C[i])  
                \oplus  \sigma_{k^{\text{write}}_{\delta(i),j}}(C'[i]) 
                \Bigr) 
\end{align*}
Finally, the receiver receives the message as well as the transmitted and updated authentication tag $t$.
Based on the received data, it can then compute $t^{*}$ as follows:
\begin{equation*}
        t^{*} = \bigoplus_{0\leq i<n} \left(  \sigma_{ \varphi( k^{\text{read}}_{\delta(i),e-1})}(C[i]) \oplus \sigma_{\varphi( k^{\text{write}}_{\delta(i),e-1})}(C[i]) \right) 
\end{equation*}
If no unauthorized manipulation of transmitted data took place, $t$ and $t^{*}$ are identical, and thus the integrity of the message is verified successfully.
Otherwise, at least one communication entity has been served with unauthentic data.  

\subsection{Self-Verifying Middlebox}
\label{sec:design:verify}

By default, \name{} operates in the most resource-conscious mode where only the final receiver verifies a message.
Notably, this verification not only covers the authenticity of the message as received at the final destination but also ensures the authenticity of the message as received by each on-path middlebox (with read or write permission).
However, further efforts are required to ensure on-path authenticity if the processing of messages by middleboxes causes \emph{side effects}, \ie has influences beyond the current message, \eg the injection of a control command.

Here, a middlebox cannot always exclusively rely on the final receiver to authenticate a message.
In some cases, \emph{optimistic processing}~\cite{2022_wagner_spmac, 2023_zhang_zombie} (\ie the idea of processing a likely genuine message with the knowledge that maliciousness is guaranteed to be detected within a short time span) still allows offloading authentication to the final receiver for reliable connections (\eg TCP).
But at the very least, middleboxes with side effects that communicate over lossy channels must authenticate the data they process themselves.
Otherwise, an attacker could manipulate a message before a middlebox processes it and prevent it from being received at its final destination, \eg by jamming a wireless channel.

In cases where \emph{immediate} verification of authenticity is required (\eg for irreversible critical decisions), \name{} enables such middleboxes to \emph{self-verify} authenticity by specifically adding an additional 16-byte authentication tag $t_j$ for the middlebox $j$.
This tag $t_j$ is computed over the subset of partial tags $\sigma$ relevant to the middlebox's access rights, such that no additional cryptographic processing is necessary.
Formally, it is computed as follows:
\begin{equation*}
        \centering
        t_j = \bigoplus_{i\in \mathbb{S}^\text{read}_{j} \cup \mathbb{S}^\text{write}_{j} } \left(  
        \sigma_{k^{\text{read}}_{\delta(i),0}}(C[i]) \right) \oplus \bigoplus_{i\in \mathbb{S}^\text{write}_{j} } \left(  
        \sigma_{k^{\text{write}}_{\delta(i),0}}(C[i])  \right) 
\end{equation*}
Like the main authentication tag $t$, these tags $t_j$ are modified by preceeding middleboxes.
These modifications are, however, restricted to the subset of contexts both middleboxes have access to, which is communicated to the middleboxes during the handshake. 
For write access, a middlebox $k$ would modify the tag $t_j$ as follows:
\begin{align*}
                t_j \overset{\oplus}{=} \bigoplus_{i\in \mathbb{S}^\text{write}_{j} \cap \mathbb{S}^\text{write}_{k} } 
                \Bigl( & \sigma_{ \varphi( k^{\text{read}}_{\delta(i),k}    )}(C[i])
                \oplus \sigma_{k^{\text{read}}_{\delta(i),k}}(C'[i])  \oplus \phantom{} \\
                & \sigma_{\varphi(k^{\text{write}}_{\delta(i),k})}(C[i])  
                \oplus  \sigma_{k^{\text{write}}_{\delta(i),k}}(C'[i]) 
                \Bigr)
\end{align*}
Analogously, the intermediary middlebox $k$ only updates the read tags if it and middlebox $j$ have access to a context:
\begin{align*}
        t_j \overset{\oplus}{=} \bigoplus_{i\in \mathbb{S}^\text{read}_{j} \cap (\mathbb{S}^\text{read}_{k} \cup \mathbb{S}^\text{write}_{k}) } 
        \Bigl( & \sigma_{ \varphi( k^{\text{read}}_{\delta(i),k}    )}(C[i])
        \oplus \sigma_{k^{\text{read}}_{\delta(i),k}}(C'[i])
        \Bigr)
\end{align*}
Finally, upon reception of the message by middlebox $j$, this middlebox can verify the authenticity of the data it has access to by recomputing $t_j$ and comparing it to the transmitted tag.
When forwarding the message to the final receiver, middlebox $j$ removes the tag $t_j$ as it is no longer needed.
After learning about self-verifying middleboxes, we can now look at how \name enables limited message injection as required by some industrial use cases.

\subsection{Limited Message Injection Capabilities}
\label{sec:design:inject}

So far, \name allows authorized middleboxes to read and write to well-defined segments of transmitted messages.
Still, certain industrial use cases for middleboxes additionally require capabilities to actively inject new messages, \eg to enable low latency control of robot arms~(cf.\ Section~\ref{sec:background:usecases}).
However, giving such middleboxes the ability to inject \emph{arbitrary} messages breaks the least-privilege principle as a middlebox could take control beyond what is necessary for its dedicated task.
Consequently, \name needs to enforce \emph{limited injection capabilities} where middleboxes can only inject those messages relevant to its intended functionality.

\name{} realizes such limited injection capabilities through the use of pre-defined \emph{message templates}.
In essence, a template is a message with dedicated placeholders that is authenticated by the endpoint and sent to the middlebox in advance.
The middlebox then has restricted write access to the placeholder segments of this message before it is forwarded~(\eg to only be able to transmit specific control commands).


However, DTLS uses nonces that are explicitly transmitted to prevent replay attacks.
Either way, for messages generated via message templates, we must ensure that \textit{(1)} the used nonce is unique and \textit{(2)} the receiver knows and accepts the nonce.
In DTLS, nonces are a combination of a sequence number and an epoch.
To not interfere with this procedure, \name defines a new content type (\texttt{0x1F}) to mark middlebox-injected messages.
These messages use the same encryption and authentication keys as normal messages but start at any unique epoch in the future.
This epoch should be chosen according to how many injected messages are expected in relation to normal messages and how many message-injecting middleboxes exist in a given communication session.

While a middlebox with the mere task of sending emergency stops in critical situations only needs the ability to send a few messages, a middlebox responsible for low-latency control adjustments continuously injects a significant number of messages.
As the sequence numbers of injected and regularly transmitted messages (from the original sender) are decoupled, the impersonated endpoint can asynchronously provide multiple authentication tags in advance for increasing sequence numbers of the same message template without having to retransmit the template itself. 
Hence, \name achieves the same fine-grained access control over injected messages as for read and write access to existing messages.

\section{The \name Handshake Protocol}%
\label{sec:handshake}

\begin{figure*}
        \includegraphics[]{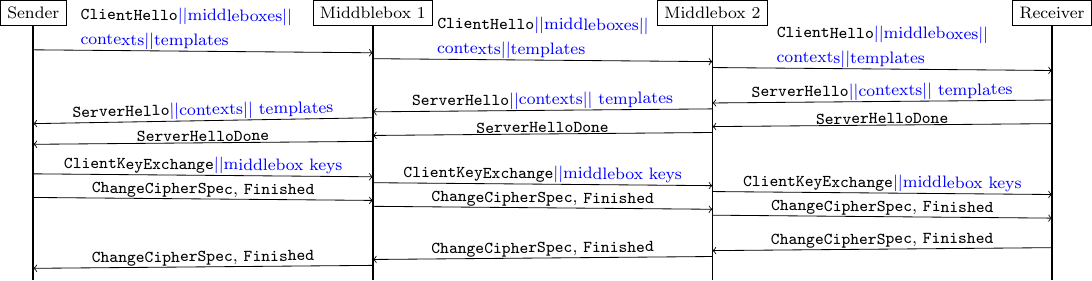}
        \caption{For \name, we extend the DTLS 1.2 handshake and add additional data to select messages (highlighted in blue).}
        \label{fig:handshake}
\end{figure*}

\name requires several keys to enable fine-grained access control for middleboxes.
These keys can be pre-configured or distributed ad-hoc by a trusted party.
In many cases, it is, however, beneficial if the involved parties can agree upon these keys by themselves.
Therefore, we adapt the DTLS 1.2 handshake to additionally exchange the keys required for \name to the endpoints as well as all involved middleboxes.
For simplicity, we assume that all communication entities have pre-established shared secrets, which is often reasonable as all entities are known in advanace and managed by a single operator in industrial networks.
To enable certificate-based authentication of all entities, we can employ the \texttt{ServerKeyExchange} message proposed in step 3 of the mcTLS handshake~\cite{2015_naylor_mctls}, which we, however, do not require due to our pre-shared keys.
Figure~\ref{fig:handshake} highlights the necessary additions to the DTLS 1.2 handshake in blue.
In the following, we discuss these changes step by step.

\textbf{\texttt{ClientHello.}}
The \texttt{ClientHello} announces the desire to establish a new communication session and proposes a set of cipher suites.
The \texttt{ClientHello} contains a pre-exchanged cookie to thwart \ac{DoS} attacks as per the DTLS 1.2 standard, which we do not change.
One important change is that the \texttt{ClientHello} message passes through all middleboxes, which are thus informed about the new session and may remove cipher suites from the announced list.
We extend it by the information concerning middleboxes and their access rights.
As additional information, \name first appends a list of middleboxes identified by their IP address.
Then, the \texttt{contexts} field defines all contexts needed by that session, \ie combinations of read and write access for the different middleboxes.
Finally, the client appends a list of possible message templates.
First, the number of templates is announced, followed by a null-byte terminated enumeration of \texttt{segmentation info} fields, as introduced in Figure~\ref{fig:template}.
If the receiver accepts the request and supports a requested cipher suite, it responds with a \texttt{ServerHello} message.

\textbf{\texttt{ServerHello.}} The receiver responds with a \texttt{ServerHello} handshake message, agrees on a selected cipher suite, and replays the \name-specific \emph{contexts} and \emph{template} fields.
These fields must be replayed as the server may add more contexts or templates to it.

\textbf{\texttt{ServerHelloDone}.} The \texttt{ServerHelloDone} does not require any modification as it only signals the end of the \texttt{ServerHello}.

\textbf{\texttt{ClientKeyExchange}.}
The client then shares the final information the server needs to derive the symmetric keys used in this session and shares all middlebox keys $k^{\text{read}}$ and $k^{\text{write}}$ with the respective involved middleboxes through a \texttt{ClientKeyExchange} message.
For this key distribution, the sender first computes a key to encrypt and authenticate data to the respective middleboxes.
Based on $\text{secret}_{s,m}$ shared between the sender $s$ and a middlebox $m$, these key distribution keys $k^{kd}$ are derived as:
\[
    k^\mathit{kd}_m = \textsc{kdf} (\mathrm{secret}_{s, m}, \mathrm{nonce}),
\]
where the nonce is derived from the standard DTLS handshake.
Then, the context encryption keys $k^\text{enc}$ are derived independently of any secrets shared with a middlebox, as 
\[
        \textsc{kdf}(\mathrm{secret}{s,r}, \mathrm{nonce}, \mathrm{context}, \text{`encrypt'})
\]
from the secret shared between sender $s$ and receiver $r$, the nonce, an identifier of the respective context and a unique string.
Finally, the context authentication keys $k^\text{read}$ and $k^\text{write}$ are derived as
\[
        \textsc{kdf}(\mathrm{secret}_{s,r}, \mathrm{nonce}, \mathrm{middlebox}, \mathrm{context}, \{\text{`read'},\text{`write'}\})
\]
by including an identifier of the targeted middlebox and different strings for read or write keys.
To ensure that a middlebox only gains access to its keys (and the respective preceding keys as given by $\varphi(\cdot)$), the client encrypts each middlebox's context keys with the respective key distribution key $k^\mathit{kd}_m$ before appending them to the \texttt{ClientKeyExchange} message.
Thus, if middlebox~2 has read access to context~1 and write access to context~3, the sender appends $\text{enc}_{k^{kd}_{s,2}}(k_{2,1}^{\text{read}}||k_{2,3}^{\text{read}}||k_{2,3}^{\text{write}}||\varphi (k_{2,1}^{\text{read}})||\varphi (k_{2,3}^{\text{read}})||\varphi (k_{2,3}^{\text{write}}))$ to the \texttt{ClientKeyExchange} message.
The middleboxes can then decrypt these respective keys as the \texttt{ClientKeyExchange} passes.
The subsequently transmitted \texttt{ChangeCipherSpec} and \texttt{Finished} messages are not changed for \name.

\textbf{Final Server Messages.}
In DTLS 1.2, the server sends a final copy of \texttt{ChangeCipherSpec} and \texttt{Finished} messages to conclude the handshake.
This procedure is not changed by \name as the receiver can compute and verify all key distribution keys that the sender transmitted in the \texttt{ClientKeyExchange} message.
Overall, a \name session can be efficiently established, and we now look at its performance in real-world scenarios.

\section{Performance Evaluation}
\label{sec:evaluation}

\name fulfills all functional requirements expected of a middlebox-aware security protocol for industrial networks~(\cf~Section~\ref{sec:motivation:requirements}).
To evaluate if its performance is suitable even for resource-constrained devices, we implemented a prototype for Contiki-NG 4.8, a popular operating system for IoT devices, by extending the \emph{tinydtls} library.

\subsection{\name vs. the Current State-of-the-Art}

First, we compare the processing latency of \name to related approaches.
These approaches consist of the tinydtls DTLS 1.2 implementation (which offers no middlebox awareness) and a custom mcTLS implementation\footnote{We could not source any implementations of the approaches discussed in Section~\ref{sec:background:related_work} and therefore implement the closest competitor of \name ourselves.} adapted to DTLS to avoid TCP overhead.
For all protocols, we send payloads of lengths increasing from 1 to \SI{256}{bytes}.
\name uses a single write context to emulate the same functionality as mcTLS.
Our measurements run on a Zolertia RE-Mote~(Cortex\,M3\,@\,32\,MHz, 32-bit\,CPU), a common device to represent the constraints of industrial hardware~\cite{2019_righetti_performance,2022_wagner_bpmac,2022_wagner_spmac}.

\begin{figure}
        \includegraphics[width=\columnwidth]{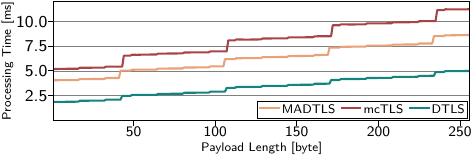}
        \caption{\name performs better than mcTLS in comparable scenarios, where middlebox write access is given for selected messages. DTLS provides no such guarantees.}
        \label{fig:protocol_comparision}
\end{figure}

Figure~\ref{fig:protocol_comparision} plots the processing time against the length of the transmitted payload for the different protocols.
All protocols require more processing as the payload increases.
However, these increases are discrete, marginally rising by about \SI{0.1}{\milli\second} whenever the payload fills up a new AES block (all \SI{16}{bytes}), and more substantially jumping by about \SI{0.4}{\milli\second} when sha256 blocks of the HMAC computation are filled up~(all \SI{64}{bytes}).
Thus, \name  is efficient even for large messages and even achieves a noticable performance gain against mcTLS which neither offers the same fine-grained data access as \name nor ensures path integrity.

Beyond latency, reducing bandwidth usage is also imperative for \name.
Here, the DTLS~1.2 record protocol carries a header overhead (including \SI{16}{bytes} tags) of \SI{30}{bytes} while mcTLS's overhead is \SI{63}{bytes}.
Meanwhile, \name only adds \SI{1}{byte} to DTLS~1.2 for the \texttt{template id} that defines the message structure.
Even for multiple contexts, \name does not require more space.

\name is thus attractive performance-wise for industrial networks, even if no fine-grained data access is needed.
In the case of a single context, it saves over \SI{22}{\%} of processing latency (more for larger messages) and \SI{32}{bytes} of header over mcTLS.
Most importantly, \name offers this performance while offering true least-privilege data access to middleboxes, ensuring path integrity, and allowing least privilege traffic injection, three crucial features not offered by mcTLS, or any other approach from related work.

\textbf{Cipher Suite for Faster Processing.}
\name performs adequately for many industrial applications.
However, some applications require even faster processing in the sub-millisecond range.
To assess \name under these harsh requirements, we design a cipher suite to minimize processing latency based on recent work on preprocessed encryption and integrity protection.
More concretely, we create a cipher suite based on antedated encryption~\cite{2018_hiller_secure} and BP-MAC~\cite{2022_wagner_bpmac}.
Antedated encryption uses AES in counter mode and splits it into a precomputation phase for the computationally-intensive keystream generation and a fast online phase that only XORs a message with the cached keystream~\cite{2018_hiller_secure}.
BP-MAC, in turn, achieves fast integrity protection for short messages by combining precomputed authentication tags for individual bits via a Carter-Wegman construction~\cite{2022_wagner_bpmac}.
Using this cipher suite for \name on our resource-constrained RE-Mote board, message encryption and authentication time reduces to \SI{624}{\micro\second} (compared to \SI{3.975}{\milli\second}) for a 5-byte message with a single write context spanning the entire message.
Thus, with suitable cryptographic algorithms, \name is apt for low-latency scenarios on constrained hardware.

\subsection{Impact of the Size and Number of Contexts}

\begin{figure}
        \includegraphics[width=\columnwidth]{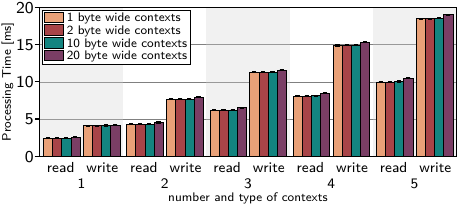}
        \caption{\name’s performance scales linearly in the number of contexts, while their sizes only have marginal impact.}
        \label{fig:contexts}
\end{figure}

We must also understand the impact of \name' fine-grained access control on its performance. 
Therefore, we evaluate the impact of the number and size of contexts on the processing time.
We continue using the Zolertia RE-Mote as evaluation platform and measure the encryption and authentication time of a 100-byte long packet.
We add 1 to 5 contexts of sizes varying between 1 and 20 bytes to each message.
We repeat our measurements 20 times and show the results, including 99\%-confidence intervals, in Figure~\ref{fig:contexts}.

We observe a linear growth of processing times with the number of contexts and that write contexts require more processing than read contexts.
This scaling is expected as it is proportional to the number of calls to HMAC-SHA256.
While processing one 20-byte read context takes \SI{2.42}{\milli\second}, the same operation over a write context lasts \SI{4.22}{\milli\second}.
This behavior can also be explained by the number of HMAC-SHA256 calls, as write contexts require two calls per context compared to the one required for read contexts.
Meanwhile, the size of the contexts only has a negligible impact on the processing time as all contexts individually fit into a single SHA256 block.

Overall, \name performs adequately even for scenarios requiring many contexts for different middleboxes with diverse functionalities.
As contexts are shared if the same data is accessed by multiple middleboxes, even complex chains of middleboxes can operate in a least-privilege mode with a low number of contexts.
Thus, even resource-constrained devices, commonly seen as endpoints in industrial networks, can employ \name.

\subsection{\name Across Different Hardware Classes}

Middleboxes are typically more powerful as they have to process many messages from different sources.
Therefore, we use the Raspberry Pi Zero~(which uses the same Arm 11 chip that the Netronome NFP-4000 Flow Processor uses for general-purpose processing) to evaluate the performance of SmartNIC-based middleboxes without further optimizations.
Moreover, we evaluate \name on an AMD EPYC 7551 server CPU.
Thus, we can learn how \name performs over a wide variety of CPU classes and architectures.

\begin{figure}
        \includegraphics[width=\columnwidth]{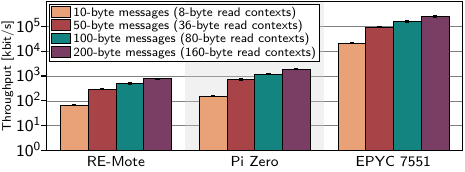}
        \caption{Throughput is limited on constrained devices but middleboxes can take advantage of more powerful hardware.}
        \label{fig:hardware}
\end{figure}

For our evaluation, we process 10, 50, 100, and 200 byte messages with the middlebox accessing \SI{80}{\percent} of a message's data via a single read context.
The processing entails the decryption of the accessed data and updating the authentication tag. 
We repeat all measurements 20 times and report on the throughput with 99\%-confidence intervals in Figure~\ref{fig:hardware}.
Across all processors, the throughput grows significantly with longer messages as the fixed per-message overhead mainly impacts small messages.

Our evaluation shows that if devices like the Zolertia RE-Mote were used as a middlebox, they would still achieve a throughput of over \SI{66}{kbit/s}.
The Raspberry Pi Zero more than doubles this throughput, with a throughput between 154 and \SI{1877}{kbit/s}, depending on the message sizes.
Expectedly, SmartNICs (as well as programmable switches) can, however, not rely on a relatively slow general-purpose processing core to achieve gigabit throughput.
Still, their performance can be significantly optimized with device-specific implementations of cryptographic primitives~\cite{2022_kottur_implementing, 2021_yoo_secure}.
On the other hand, our AMD EPYC 7551 processor achieves a throughput between 20 and \SI{249}{Mbits/s} on a single of its 32 cores.

\subsection{\name in the Real World}

To verify \name's utility in real-world use cases, we conduct two case studies.
First, we implement a middlebox that translates between local coordinates of robot arms and global coordinates as proposed by Kunze~\etal~\cite{2021_kunze_investigating}.
Here, \name allows us to ensure that the middlebox only accesses the first six byte of a packet's payload where the x, y, and z coordinates are encoded in 16 bits each.
The remaining payload, containing supplementary sensor readings (\eg timestamp, grip pressure, gripper rotation) in an additional 14 bytes, cannot be written to or read by the middlebox.
This data layout, \eg corresponds to one observed in Modbus communications.
Secondly, we use \name to realize an IDS based on Snort rules.
To this end, we use the 14 Quickdraw Snort rules~\cite{rules} for industrial Modbus communication.
Using \name{}, we constrain the IDS' access to selectively reading 3, 5, or 6 bytes per Modbus frame, depending on the communication flow.
Without restricting the IDS' capabilities, \name thus blinds over \SI{60}{\percent} of all bytes in the corresponding Modbus test traces~\cite{rules}.

\section{Limitations of \name}

\name addresses many shortcomings of related work on secure middlebox-aware (industrial) communication.
These achievements come, however, with a few drawbacks that must be considered before deploying \name.
First, \name' handshake is significantly more complex than the standard DTLS~1.2 handshake.
While it does not require additional round trips, more data must be exchanged between the entities.
Per se, this is not a problem in industrial networks with relatively static connections since handshakes can be performed in advance during non-critical periods.
However, the added complexity makes weaknesses in design and implementation more likely and may be restrictive in some scenarios.

Secondly, \name needs more extensive key management and storage than simple end-to-end communication. 
While this drawback is inevitable when multiple entities with different access rights on a single communication channel are involved, it may impact embedded devices with limited storage capabilities.
Also, the corresponding key exchange protocol, as exemplified in Section~\ref{sec:handshake}, becomes more error-prone in terms of design and implementation.

Thirdly, \name offloads integrity verification to the final receiver of a message per default. 
However, it must be carefully considered whether middleboxes must additionally verify the integrity of processed data themselves.
Still, \name offers efficiently computed additional middlebox-specific authentication tags that require no additional cryptographic processing to append to a message.

Fourthly, while \name is designed with performance in mind and even outperforms its closest competitor (mcTLS~\cite{2015_naylor_mctls}), the added processing overhead is not negligible for resource-constrained devices in industrial scenarios.
Fortunately, \name's processing adds minimal jitter, such that a deterministic overhead can be considered when designing control algorithms where required~\cite{2018_ruth_towards}.

Fifthly, \name adopts DTLS 1.2 with our building blocks to bring middlebox-aware security to industrial communication.
In reality, the industrial landscape and beyond also uses other security protocols (\eg TLS) that can and should not always be replaced with DTLS.
While we see nothing preventing the adoption of other protocols with \name's features, a concrete design must still be proposed to bring our advances to a wider variety of applications.

Sixthly, \name is most efficient for predictable message structures as they are often found in the \ac*{IIoT}.
Employing \name on the traditional Internet would expose it to more dynamic content (\eg websites).
While the explicit transmission of segmentation info enables such scenarios, the added bandwidth overhead must be considered.
Moreover, privacy implications of metadata (\eg for templates) must be carefully considered in other scenarios, as devices no longer belong to a single operator in that case.

\name thus mostly exhibits limitations outside the industrial domain.
However, domain-specific adaptions can alleviate some of these constraints while benefiting from \name's main contributions: Allowing least-privilege access control on a communication channel for middlebox processing.

\section{Conclusion}

This paper proposes \name, a middlebox-aware enhancement of the DTLS protocol tailored explicitly to the \ac{IIoT}.
Hereby, \name addresses multiple major limitations of the current state-of-the-art on middlebox-aware security that focuses heavily on the traditional Internet.
Most importantly, \name allows fine-grained access control for middleboxes on a bit-level and enables middleboxes to inject a restricted set of messages where this is desired (\eg for emergency shutdowns) while still operating more efficiently than mcTLS~\cite{2015_naylor_mctls}.

Specifically, \name segments messages and assigns contexts (\ie read and write access rights) according to the middleboxes' needs.
Each segment is encrypted and authenticated separately without expanding the packet.
Middleboxes are permitted to read or write to specific segments and either verify integrity directly or defer this step to the final receiver for efficiency reasons.
\name processes packets in only a few milliseconds on heavily constrained hardware, while performance scales linearly with the number of contexts and message lengths.
Meanwhile, \name achieves up to \SI{249}{Mbit/s} throughput on a single AMD EPYC 7551 core, enabling middleboxes to process many different communication streams.

\name thus brings middlebox-aware security to the \ac{IIoT} to solve the dilemma where middleboxes become increasingly popular and thus prevent secure end-to-end communication.

\begin{acks}
We would like to thank Ren\'e Glebke and our anonymous reviewers for their valuable feedback.
This paper was supported by the EDA Cyber R\&T project CERERE, funded by Italy and Germany.
Funded by the Deutsche Forschungsgemeinschaft (DFG, German Research Foundation) under Germany's Excellence Strategy – EXC-2023 Internet of Production – 390621612.
The authors are responsible for the contents of this work.
\end{acks}

\bibliographystyle{ACM-Reference-Format}
\bibliography{paper}

\appendix

\section*{Appendix}
\section{Soundness of \name}
\label{appendix:soundness}

We first prove the soundness of \name's authentication scheme, \ie that without malicious manipulation all valid tags are accepted.
First, we look at the case when no middlebox is included.

\begin{equation*}
        \centering
        \begin{aligned}
                t^{*} & = \bigoplus_{0\leq i<n} \left(  \sigma_{ \varphi( k^{\text{read}}_{\delta(i),j-1})}(X[i]) \oplus \sigma_{\varphi( k^{\text{write}}_{\delta(i),j-1})}(X[i]) \right) \\
                & =  \bigoplus_{0\leq i<n} \left(  \sigma_{ k^{\text{read}}_{\delta(i),0}}(X[i]) \oplus \sigma_{ k^{\text{write}}_{\delta(i),0}}(X[i]) \right) \\
                & \overset{!}{=} t
        \end{aligned}
\end{equation*}

Then, we prove by induction that the inclusion of a reading middlebox $k$ as last hop before the intended receiver does transform a valid tag $t'$ into a valid final tag $t^{*}$.

\begin{equation*}
        \centering
        \begin{aligned}
                t' & = && \bigoplus_{0\leq i<n} \left(  \sigma_{ \varphi( k^{\text{read}}_{\delta(i),k})}(X[i]) \oplus \sigma_{\varphi( k^{\text{write}}_{\delta(i),k})}(X[i]) \right) \\
                & \overset{ \text{read}}{=} &&\bigoplus_{0\leq i<n} \left(  \sigma_{ \varphi( k^{\text{read}}_{\delta(i),k})}(X[i]) \oplus \sigma_{\varphi( k^{\text{write}}_{\delta(i),k})}(X[i]) \right) \oplus \\ 
                &  && \bigoplus_{i\in \mathbb{S}^\text{read}_{k} } \left(  
                \sigma_{ \varphi( k^{\text{read}}_{\delta(i),k}    )}(X[i]) 
                \oplus \sigma_{k^{\text{read}}_{\delta(i),k}}(X[i])  \right) \\
                & = &&\bigoplus_{i \notin \mathbb{S}^\text{read}_{k}} \left(  \sigma_{ \varphi( k^{\text{read}}_{\delta(i),k})}(X[i]) \oplus \sigma_{\varphi( k^{\text{write}}_{\delta(i),k})}(X[i]) \right) \oplus \\ 
                &  && \bigoplus_{i\in \mathbb{S}^\text{read}_{k} } \left( \sigma_{k^{\text{read}}_{\delta(i),k}}(X[i]) \oplus \sigma_{\varphi( k^{\text{write}}_{\delta(i),k})}(X[i])  \right) \\
                & =   && \bigoplus_{0\leq i<n} \left(  \sigma_{ \varphi( k^{\text{read}}_{\delta(i),j-1})}(X[i]) \oplus \sigma_{\varphi( k^{\text{write}}_{\delta(i),j-1})}(X[i]) \right) \\
                & \overset{!}{=} t^{*}
        \end{aligned}
\end{equation*}

Finally, we do the same inductive proof for a writing middlebox.

\begin{equation*}
        \centering
        \begin{aligned}
                t' & = && \bigoplus_{0\leq i<n} && \left(  \sigma_{ \varphi( k^{\text{read}}_{\delta(i),k})}(X[i]) \oplus \sigma_{\varphi( k^{\text{write}}_{\delta(i),k})}(X[i]) \right) \\
                & \overset{ \text{write}}{=} &&\bigoplus_{0\leq i<n} &&\left( \sigma_{ \varphi( k^{\text{read}}_{\delta(i),k})}(X[i]) \oplus \sigma_{\varphi( k^{\text{write}}_{\delta(i),k})}(X[i]) \right) \oplus \\ 
                &  && \bigoplus_{i\in \mathbb{S}^\text{write}_{j} } && \left(  \sigma_{ \varphi( k^{\text{read}}_{\delta(i),j} )}(X[i]) \right.
                \oplus \sigma_{k^{\text{read}}_{\delta(i),j}}(X'[i])  \oplus \\
                & && && \phantom{\Bigl(\,} \sigma_{\varphi(k^{\text{write}}_{\delta(i),j})}(X[i])  \left. \oplus  \sigma_{k^{\text{write}}_{\delta(i),j}}(X'[i]) \right)    \\
                & = &&\bigoplus_{i \notin \mathbb{S}^\text{write}_{k}} &&\left( \sigma_{ \varphi( k^{\text{read}}_{\delta(i),k})}(X'[i]) \oplus \sigma_{\varphi( k^{\text{write}}_{\delta(i),k})}(X'[i]) \right) \oplus \\ 
                &  && \bigoplus_{i\in \mathbb{S}^\text{write}_{k} } && \left( \sigma_{k^{\text{read}}_{\delta(i),k}}(X'[i]) \oplus \sigma_{ k^{\text{write}}_{\delta(i),k}}(X'[i])  \right) \\
                & =   && \bigoplus_{0\leq i<n} && \left(  \sigma_{ \varphi( k^{\text{read}}_{\delta(i),j-1})}(X'[i]) \oplus \sigma_{\varphi( k^{\text{write}}_{\delta(i),j-1})}(X'[i]) \right) \\
                & \overset{!}{=} t^{*}
        \end{aligned}
\end{equation*}

\section{Security of \name}
\label{appendix:security}

Besides the soundness of \name's authentication scheme, its security is also crucial.
To prove its security, we first look at the security definition of traditional \ac{MAC} schemes to show that \name cannot be attacked by outsiders.
Afterwards, we also prove that an insider, \ie a middlebox authorized to read or write some part of a packet, cannot manipulate a packet beyond what it is authorized to do.

\subsection{Security of Traditional MAC Schemes}
The security of a MAC scheme $\Sigma = (\text{Sig}, \text{Vrfy})$ over $(\mathcal{K},\mathcal{M},\mathcal{T})$ is typically defined through a game between a challenger and an adversary $\mathcal{A}$~\cite{2020_boneh_crypto}.

\textbf{Attack Game 1.}
\begin{itemize}
        \item The challenger randomly picks a key $k$ from $\mathcal{K}$.
        \item The adversary queries an oracle with a message $m_i$ for a valid tag $t_i$, \ie $t_i$ such that $\text{Vrfy}_k(m_i, t_i)$ returns \texttt{accept}. Denote $\mathbb{M}$ the set of all $m_i$ queried by $\mathcal{A}$.
        \item Eventually, $\mathcal{A}$ outputs a candidate forgery $(m,t) \in \mathbb{M} \times \mathcal{T}$, with $m \notin \mathbb{M}$.  
\end{itemize}

$\mathcal{A}$ wins the above game if $\text{Vrfy}_k(m, t)$ returns \texttt{accept}. 
We define $\mathcal{A}$'s advantage with respect to $\Sigma$, denoted as $\text{Adv}_\mathcal{A} [ \Sigma ]$, as the probability that $\mathcal{A}$ wins the game. 
A MAC scheme is considered secure if the advantage of all efficient adversaries $\mathcal{A}$ with respect to $\Sigma$ is negligable, \ie  $\text{Adv}_\mathcal{A}[\Sigma] = Pr[\mathcal{A} \text{ wins Attack Game 1 for }\Sigma]  < \epsilon$.

\subsection{\name is Secure Against Outsiders}

Our proof is built upon the fact that $\text{Adv}_\mathcal{B}[\Sigma^{\oplus}]  \leq \text{Adv}_\mathcal{A}[\Sigma]$, where a tag $t$ by $\Sigma^{\oplus}$ is computed as $t = \Sigma(X[1])\oplus \dots \oplus \Sigma(X[n])$, with $m=X[1]||\dots||X[n]$.
We first adapt Attack Game 1 to prove the security of a MAC scheme that allows the manipulation (of selected parts) of the message by authorized middleboxes.\\

\textbf{Attack Game 2.}
\begin{itemize}
        \item The challenger randomly picks a key $k$ from $\mathcal{K}$.
        \item The adversary queries an oracle with a segment $X_i[\cdot]$ for a valid partial tag $t_i[\cdot]$, \ie $t_i[\cdot]$ such that $\text{Vrfy}^{\Sigma}_k(X_i[\cdot], t_i[\cdot])$ returns \texttt{accept}. Denote $\mathbb{M}$ the set of all $m_i$ that can be composed of partial messages queried by $\mathcal{A}$.
        \item Eventually, $\mathcal{A}$ outputs a candidate forgery $(m,t) \in \mathcal{M} \times \mathcal{T}$, with $m \notin \mathbb{M}$.  
\end{itemize}

In Attack Game 2, the adversary is strictly more powerful than in Attack Game 1, as they can query partial authentication tags, but still (over multiple queries) learn the tag over every specific message.
Thus, \name is secure against outsider attacks if an adversary $\mathcal{A}$ cannot win Attack Game 2 with a non-negligible probability.

Still, the advantage of $\mathcal{A}$ to win Attack Game 2 for $\Sigma^{\oplus}$ remains negligible:
For every $m \notin \mathbb{M}$, there exists at least one $m_i$ for which $t_i$ was not queried by $\mathcal{A}$.
Thus, $\text{Adv}_\mathcal{B}[\Sigma^{\oplus}]$  can be at most $\text{Adv}_\mathcal{A}[\Sigma]$, as otherwise $\mathcal{A}$ could learn $t_i$ for $X_i[\cdot]$ which were not queried, \ie $\text{Adv}_\mathcal{B}[\Sigma^{\oplus}] \leq \text{Adv}_\mathcal{A}[\Sigma]$.
Consequently, \name's authentication scheme is secure as long as the underlying MAC scheme is secure and the tags $t_i$ computed for different message segments $X_i[\cdot]$ are independent.

\subsection{ \name is Secure Against Insiders} 
We established thus far that only authorized writers can alter (segments of) messages such that they are successfully verified by the receiver.
However, we have not considered the possibility of ephemeral changes by malicious actors, \ie the temporary altering of a message for only one or a few middleboxes' processing, before the message is altered back to a message accepted by the final receiver.
To validate that such an attack is not possible, we have to take a deeper look at how a message's authentication tag changes over time.

When middlebox $j$ accesses a segment, it alters the authentication tag in a deterministic way: $t \oplus \sigma_{\varphi( k_{\_,j}    )}(X[\cdot]) \oplus \sigma_{k_{\_,j}}(X[\cdot])$.
The partial tags from the last accessing entity are first removed from the aggregated tag, before new partial tags is added by the current entity.
This process might be done once or twice, depending on whether the accessing middlebox is only reading or also writing.

Consequently, if an attacker maliciously changes the message segment $X'[\cdot]$, they must be able to compute $\sigma_{\varphi( k_{\_,j} )}(X'[\cdot])$ and $\sigma_{k_{\_,j}}(X'[\cdot])$ in order to remove the consequences of their attack from the authentication tag.
It is, however, precisely the middlebox $j$ doing these computations that has access to the required keys (besides the endpoints of the communication channel).
The next middleboxes accessing that specific segment may compute $\sigma_{k_{\_,j}}(X'[\cdot])$ for the attacker if the message has not been changed back yet. 
However, then $\sigma_{\varphi^{-1}( k_{\_,j}    )}(X'[\cdot])$ needs to be intercepted by the attacker.
In the end, the attacker needs to compute two partial authentication tags to ephemerally change a message in an unauthorized way.
Thus, the collusion of at least two middleboxes having read access to a specific message segment is required for an attack that introduces ephemeral changes.
Hence, no single actor, external or internal, can attack \name's authentication scheme.

\end{document}